\documentclass[runningheads,a4paper]{llncs}
\usepackage[utf8]{inputenc}
\usepackage[T1]{fontenc}
\usepackage{graphicx}
\usepackage{longtable}
\usepackage{float}
\usepackage{wrapfig}
\usepackage{rotating}
\usepackage[normalem]{ulem}
\usepackage[fleqn]{amsmath}
\usepackage{textcomp}
\usepackage{marvosym}
\usepackage{wasysym}
\usepackage{amssymb}
\usepackage{amsthm}
\usepackage{hyperref}
\usepackage{listings}
\usepackage{sty/bsymb}
\usepackage{xcolor}
\usepackage{algorithm, algorithmicx, algpseudocode}
\usepackage{todonotes}
\usepackage{cite}
\author{Tsutomu Kobayashi\inst{1,2}\Letter, Rick Salay\inst{3}, Ichiro Hasuo\inst{2}, Krzysztof Czarnecki\inst{3}, Fuyuki Ishikawa\inst{2}, and Shin-ya Katsumata\inst{2}}
\urldef{\niimails}\path|{t-kobayashi, hasuo, f-ishikawa, s-katsumata}@nii.ac.jp|
\urldef{\uwmails}\path|{rsalay,kczarnec}@gsd.uwaterloo.ca|
\institute{
Japan Science and Technology Agency, Saitama, Japan \and
National Institute of Informatics, Tokyo, Japan \\ \niimails \and
University of Waterloo, Waterloo, Canada \\ \uwmails}
\title{Robustifying Controller Specifications of Cyber-Physical Systems Against Perceptual Uncertainty \thanks{The work is supported by JST ERATO HASUO Metamathematics for Systems Design Project (No. JPMJER1603) and JSPS KAKENHI grant number 19K20249. TK is supported by JST ACT-I (No. JPMJPR17UA). RS and KC are partly supported by NSERC Discovery and DND Supplement Grants.}}

\authorrunning{T. Kobayashi et al.}
\titlerunning{Robustifying Controller Specifications Against Perceptual Uncertainty}

\DeclareMathAlphabet{\mathbfit}{\encodingdefault}{\rmdefault}{bx}{it}
\DeclareMathAlphabet{\mathbfnotx}{\encodingdefault}{\rmdefault}{b}{n}
\newcommand{\tuple}[1]{\langle{#1}\rangle}

\newcommand{\plant}{\mathsf{p}}
\newcommand{\controller}{\mathsf{c}}
\newcommand{\safety}{\mathsf{S}}
\newcommand{\uncertainty}{\mathsf{U}}

\newcommand{\teq}{\mathrel{=}}
\newcommand{\tiff}{\mathrel{\iff}}

\newcommand{\stemp}{\mathit{temp}}
\newcommand{\sturn}{\mathit{tn}}
\newcommand{\hstemp}{\mathit{\widehat{temp}}}
\newcommand{\hsturn}{\mathit{\widehat{tn}}}
\newcommand{\tstemp}{\mathit{\widetilde{temp}}}
\newcommand{\pdt}{\mathit{dt}}
\newcommand{\pdc}{\mathit{dc}}
\newcommand{\pdh}{\mathit{dh}}

\newcommand{\eIdx}{\mathsf{idx}^{\controller}}
\newcommand{\ePar}{\mathsf{par}^{\controller}}
\newcommand{\eParEpsi}{\mathsf{par}^{\controller,\varepsilon}_{i}}
\newcommand{\eSafePari}{\mathsf{safpar}^{\controller,\varepsilon}_{i}}

\theoremstyle{definition}
\newtheorem{mydefinition}{Definition}
\newtheorem{myexample}{Example}

\begin{document}
 \maketitle

\begin{abstract}
Formal reasoning on the safety of controller systems interacting with plants is complex because developers need to specify behavior while taking into account perceptual uncertainty.
To address this, we propose an automated workflow that takes an Event-B model of an uncertainty-unaware controller and a specification of uncertainty as input.
First, our workflow automatically injects the uncertainty into the original model to obtain an uncertainty-aware but potentially unsafe controller.
Then, it automatically robustifies the controller so that it satisfies safety even under the uncertainty.
The case study shows how our workflow helps developers to explore multiple levels of perceptual uncertainty.
We conclude that our workflow makes design and analysis of uncertainty-aware controller systems easier and more systematic.
\end{abstract}


\section{Introduction}
The core function of controller systems is perceiving the state of the plant and taking appropriate actions to satisfy desirable properties of the plant.
In reality, however, such interactions have uncertainty.
Particularly, \emph{perceptual uncertainty}, namely the gap between the \emph{true value} of the plant and a \emph{perceived value} is significant, because basing a controller's action on an incorrect state can cause safety risk. For example, misperceiving the position of a car ahead may make the difference between a collision and safely following it \cite{salay2020purss}.
Therefore, for safety, developers need to  account for perceptual uncertainty when constructing controllers.

However, designing a controller to address its core requirements at the same time as addressing perceptual uncertainty can be complex.
In addition, the details of perceptual uncertainty may be unclear at the design phase since they can depend on the environment where the controller system is deployed.
An alternative is to add support for perceptual uncertainty to an existing controller in such a way that it provides formal safety guarantees.

\begin{figure}[tbp]
  \centering
  \includegraphics[width=.9\linewidth]{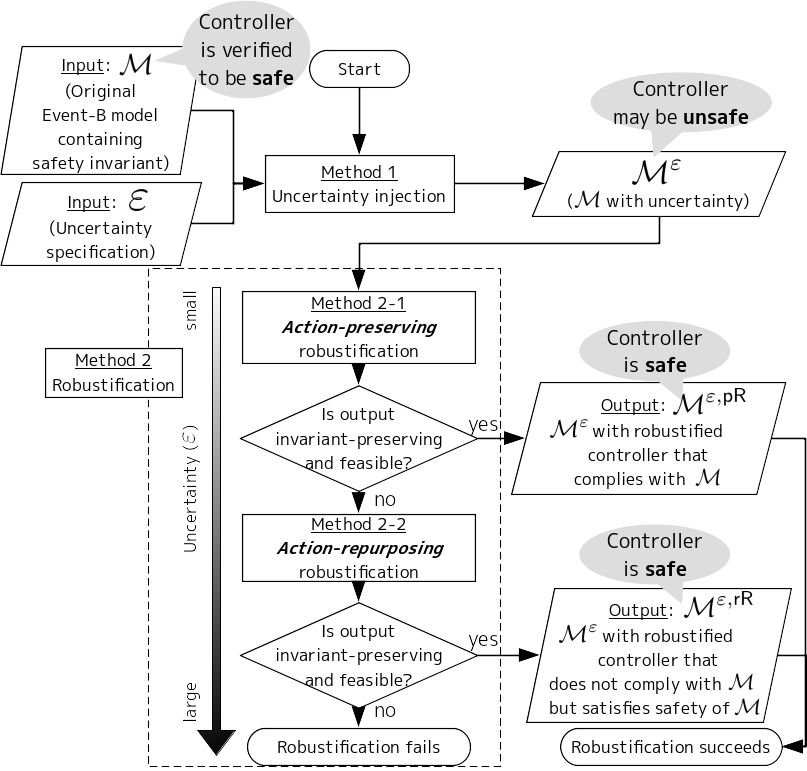}
  \caption{Overview of our uncertainty handling workflow}
  \label{fig:methods-overview}
\end{figure}

In this paper, we propose a workflow for robustifying a model of an uncertainty-unaware controller against perceptual uncertainty.
Specifically, the whole workflow (Fig.~\ref{fig:methods-overview}) is composed of three methods.
The first method (\emph{uncertainty injection}, \S~\ref{sec:injection}) takes an uncertainty-unaware model of the controller and plant (\emph{original model} $\mathcal{M}$, \S~\ref{sec:ctrlPlantModel}) and a specification of perceptual uncertainty (\emph{uncertainty specification} $\varepsilon$) as the input, and injects $\varepsilon$ into $\mathcal{M}$ to obtain an uncertainty-aware version, $\mathcal{M}^{\varepsilon}$. The model $\mathcal{M}^{\varepsilon}$ may be unsafe and the next two methods attempt to \emph{robustify} it to return it to safety.
The more conservative \emph{action-preserving robustification} method is attempted first producing model $\mathcal{M}^{\varepsilon,\mathsf{pR}}$. If this model is not feasible, the more aggressive \emph{action-repurposing robustification} method is applied to $\mathcal{M}^{\varepsilon}$ to obtain model $\mathcal{M}^{\varepsilon,\mathsf{rR}}$. When the level of uncertainty is too large, $\mathcal{M}^{\varepsilon,\mathsf{rR}}$ will too fail to be feasible. In this case, the developer may take other manual actions such as upgrading sensor devices to decrease the level of uncertainty or relax the safety invariant.

Our workflow assumes Event-B \cite{abrial2010modeling} as the modeling formalism and we have implemented the workflow as a plug-in of the IDE for Event-B (\S~\ref{sec:implementation}).

With our workflow, developers can start with constructing and analyzing controllers without considering perceptual uncertainty, and then handle the uncertainty as a second step.
Moreover, the generated model of a robustified controller is suitable for analysis because it defines a set of constraints the controller should satisfy.
For instance, if we use an uncertainty specification parameterized with the level of uncertainty, then the generated model is also parameterized with the level, and therefore it facilitates the exploration of different levels such as finding the maximum allowed uncertainty.
We demonstrate this in \S~\ref{sec:casestudies}.

\paragraph{Contributions and paper structure.}

In \S~\ref{sec:ctrlPlantModel}, we introduce a special sort of Event-B model of controller systems, assumed as input.
In \S~\ref{sec:injection}--\ref{sec:casestudies}, we describe the following contributions, before discussing related work and concluding in \S~\ref{sec:rw}--\ref{sec:rwConclusion}.
\begin{itemize}
 \item A method for injecting given perceptual uncertainty into a given model (\S~\ref{sec:injection}).
 \item Two methods for automated robustification of a controller (\S~\ref{sec:robustification}).
 \item An implementation of the whole workflow (\S~\ref{sec:implementation}).
 \item A case study of analyzing the maximum allowed level of uncertainty (\S~\ref{sec:casestudies}).
\end{itemize}

\section{Controller-Plant Models in Event-B}\label{sec:ctrlPlantModel}
We require a specific format for an input Event-B model to our workflow.

\begin{mydefinition}[A controller-plant model]\label{def:controllerPlantModel}
 A \emph{controller-plant model} $\mathcal{M}$ is an Event-B model that follows the format shown in Fig.~\ref{fig:controllerPlantModel}.
\end{mydefinition}

\begin{figure}[tbp]
 \centering
 \begin{lstlisting}
Machine $\mathcal{M}$
  State space $\mathcal{S}$
  Invariants 
    Safety invariant $I^{\safety}\subseteq \mathcal{S}$
  Initial states $A_{0}\subseteq \mathcal{S}$
  Transition function $\pi: \mathcal{S} \rightarrow \mathcal{P}(\mathcal{S})$, given by
    $\begin{array}[]{rl} \pi(s)=    & \bigcup\bigl\{\,A^{\plant}_{i}(s,p)\,\big|\,i\in[1,N_\plant], s\in \mathcal{S}, p\in P^{\plant}_{i},  G^{\plant}_{i}(s,p)\,\bigr\} \\ &\cup \bigcup\bigl\{\,A^{\controller}_{i}(s,p)\,\big|\,i\in[1,N_\controller], s\in \mathcal{S}, p\in P^{\controller}_{i}, G^{\controller}_{i}(s,p)\,\bigr\}, \end{array}$ where
    Plant event $E^{\plant}_{i}\quad (\text{where }i\in[1,N_\plant])$
      Parameter set $P^{\plant}_{i}$
      Guard $G^{\plant}_{i}\subseteq \mathcal{S} \times P^{\plant}_{i}$
      Action $A^{\plant}_{i}:\mathcal{S} \times P^{\plant}_{i} \rightarrow \mathcal{P}(\mathcal{S})$
    Controller event $E^{\controller}_{i}\quad (\text{where }i\in[1,N_\controller])$
      Parameter set $P^{\controller}_{i}$
      Guard $G^{\controller}_{i}\subseteq \mathcal{S} \times P^{\controller}_{i}$
      Action $A^{\controller}_{i}:\mathcal{S} \times P^{\controller}_{i} \rightarrow\mathcal{P}(\mathcal{S})$
  Subject to $\mathit{partitioning}$: $\forall s\in\mathcal{S}.\,\mathop{\exists !} i\in [1,N_\controller].\,\exists p\in P^{\controller}_{i}.\, G^{\controller}_{i}(s, p)$ `\label{line:CPModelPartitioning}`
 \end{lstlisting}
 \caption{A controller-plant model $\mathcal{M}$}
 \label{fig:controllerPlantModel}
\end{figure}

In essence, an Event-B model is a transition system $(\mathcal{S}, \pi\colon\mathcal{S}\to\mathcal{P}(\mathcal{S}))$ with a designated set $A_{0}$ of initial states, equipped with \emph{invariants} $I\subseteq \mathcal{S}$ that are meant to be transition-preserved (i.e., $s\in I\Longrightarrow \pi(s)\subseteq I$). (In an Event-B model, ``invariants'' are something stated as invariants and checked if they are indeed transition-preserved---see Def.~\ref{def:invariantPreservationFeasibility}.)
In Event-B, transitions are specified by \emph{events} $E_{i}$, each coming with a parameter set $P_{i}$, a guard $G_{i}$ (the transition is enabled if the guard is true), and a function $A_{i}\colon \mathcal{S}\times P_{i}\to \mathcal{P}(\mathcal{S})$ called an \emph{action}. 

Def.~\ref{def:controllerPlantModel} imposes the following additional key assumptions on Event-B models.
\begin{itemize}
 \item  Events are classified into \emph{plant events} and \emph{controller events}, since our target systems are closed-loop control systems with controllers and plants. $N_{\plant}$ and $N_{\controller}$ denote the numbers of plant and controller events, respectively.
 \item  A \emph{partitioning} requirement is imposed in Line~\ref{line:CPModelPartitioning}---it is the responsibility of the modeler to ensure that $\mathcal{M}$ satisfies this property. The requirement says that, from each state $s$, only one controller event $E^{\controller}_{i}$ is enabled.
\end{itemize}

The following ``correctness'' notions are standard in Event-B~\cite{abrial2010modeling}.
The presentation here is adapted to controller-plant models.
\begin{mydefinition}[Invariant preservation, feasibility]\label{def:invariantPreservationFeasibility}
 Let $\mathcal{M}$ be a controller-plant model presented as in Def.~\ref{def:controllerPlantModel}. 
\begin{itemize}
 \item 
 $\mathcal{M}$ is \emph{invariant-preserving} if 1) the safety invariant $I^{\safety}$ is indeed transition-preserved (i.e., $s\in I^{\safety}\Longrightarrow \pi(s)\subseteq I^{\safety}$), and 2) 
$A_{0}\subseteq I^{\safety}$.
 \item
 $\mathcal{M}$'s controller is \emph{feasible} if controller events have feasible actions, that is precisely, $\forall s\in I^{\safety}.\, \forall i\in[1,N_\controller].\, \bigl(\forall p.\, G^{\controller}_{i}(s,p) \Longrightarrow A^{\controller}_{i}(s,p)\neq\emptyset \bigr)$
\end{itemize}
\end{mydefinition}

When using a controller-plant model $\mathcal{M}$  as the input to the workflow in Fig.~\ref{fig:methods-overview}, we assume that $\mathcal{M}$ is invariant-preserving and feasible.

\begin{myexample}[The heater model $\mathcal{M}_{\mathrm{ht0}}$]
 The Event-B model $\mathcal{M}_{\mathrm{ht0}}$  in Fig.~\ref{fig:exampleControllerPlant} models a heater(-cooler) system in a pool.\footnote{For clarity, we use a different notation than Event-B's standard syntax~\cite{abrial2010modeling}.} 
 Due to an unstable water source, the pool temperature can randomly change (the plant event $E^{\plant}_{1}$).
 The system heats or cools the pool so that the temperature becomes between $30^{\circ}$C and $40^{\circ}$C.

 To state that the safety invariant should be checked only after the behavior of the controller (the heater system), the ``turn'' variable $\sturn$ indicates if the current state is after plant's ($\plant$) or controller's ($\controller$) behavior.
 There are three controller events (events $\mathsf{ctrl\_*}$).
 If the temperature is too cold (Lines~\ref{ex:ht0:e1Begin}--\ref{ex:ht0:e1End}), the controller uses the heater to increase the temperature by an appropriate amount $\pdh$.
 If the temperature is already appropriate (Lines~\ref{ex:ht0:e2Begin}--\ref{ex:ht0:e2End}), the controller changes the temperature within the safety region $[30, 40]$.
 If the temperature is too hot (Lines~\ref{ex:ht0:e3Begin}--\ref{ex:ht0:e3End}), the controller cools the water appropriately.
\end{myexample}

\begin{figure}[tbp]
 \centering
\begin{lstlisting}
Machine $\mathcal{M}_{\mathrm{ht0}}$
  State space $\mathcal{S} = \{\plant, \controller\} \times \mathbb{Z}$                /* variables $\sturn$ and $\stemp$ */
  Invariants
    Safety invariant $I^{\safety} = \{\tuple{\sturn,\stemp} \,\big|\, \sturn = \controller \Longrightarrow 30 \leq \stemp \leq 40\}$
  Initial states $A_{0}$
  Plant event $E^{\plant}_{1}$                /* $\mathsf{plant\_change\_temp}$ */
    Parameter set $P^{\plant}_{1} = \mathbb{Z}$                /* parameter $\pdt$ */
    Guard $G^{\plant}_{1}\tuple{\tuple{\sturn,\stemp},\pdt} \iff \top$
    Action $A^{\plant}_{1}\tuple{\tuple{\sturn,\stemp},\pdt} = \{\tuple{\sturn',\stemp'} \,\big|\, \sturn' = \plant \wedge \stemp' = \stemp + \pdt\}$
  Controller event $E^{\controller}_{1}$                /* $\mathsf{ctrl\_heat}$ */`\label{ex:ht0:e1Begin}`
    Parameter set $P^{\controller}_{1} = \mathbb{Z}$                /* parameter $\pdh$ */
    Guard $G^{\controller}_{1}\tuple{\tuple{\sturn,\stemp},\pdh} \iff \stemp < 30 \wedge 30 \leq \stemp + \pdh \leq 40$
    Action $A^{\controller}_{1}\tuple{\tuple{\sturn,\stemp},\pdh} = \{\tuple{\sturn',\stemp'} \,\big|\, \sturn' = \controller \wedge \stemp' = \stemp + \pdh\}$`\label{ex:ht0:e1End}`
  Controller event $E^{\controller}_{2}$                /* $\mathsf{ctrl\_keep\_safe}$ */`\label{ex:ht0:e2Begin}`
    Parameter set $P^{\controller}_{2} = \mathbb{Z}$                /* parameter $\pdt$ */
    Guard $G^{\controller}_{2}\tuple{\tuple{\sturn,\stemp},\pdt} \iff 30 \leq \stemp \leq 40$ $\wedge 30 \leq \stemp + \pdt \leq 40$
    Action $A^{\controller}_{2}\tuple{\tuple{\sturn,\stemp},\pdt} = \{\tuple{\sturn',\stemp'} \,\big|\, \sturn' = \controller \wedge \stemp' = \stemp + \pdt\}$`\label{ex:ht0:e2End}`
  Controller event $E^{\controller}_{3}$                /* $\mathsf{ctrl\_cool}$ */`\label{ex:ht0:e3Begin}`
    Parameter set $P^{\controller}_{3} = \mathbb{Z}$                /* parameter $\pdc$ */
    Guard $G^{\controller}_{3}\tuple{\tuple{\sturn,\stemp},\pdc} \iff 40 < \stemp \wedge 30 \leq \stemp - \pdc \leq 40$
    Action $A^{\controller}_{3}\tuple{\tuple{\sturn,\stemp},\pdc} = \{\tuple{\sturn',\stemp'} \,\big|\, \sturn' = \controller \wedge \stemp' = \stemp - \pdc\}$`\label{ex:ht0:e3End}`
 \end{lstlisting}
 \caption{The heater model $\mathcal{M}_{\mathrm{ht0}}$}
 \label{fig:exampleControllerPlant}
\end{figure}

\section{Uncertainty Injection}
\label{sec:injection}
The first step of 
 our workflow (Fig.~\ref{fig:methods-overview}) is to inject specification of potential perceptual uncertainty to  an input model $\mathcal{M}$---a  controller-plant model that does not include perceptual uncertainty. 
In the following definition, the function $\varepsilon\colon \mathcal{S}\to\mathcal{P}(\mathcal{S})$ specifies the kind of uncertainty to be taken into account.

\begin{mydefinition}[Uncertainty injection $(\underline{\phantom{x}})^{\varepsilon}$]
 Let $\mathcal{M}$ be a controller-plant model (Def.~\ref{def:controllerPlantModel}, Fig.~\ref{fig:controllerPlantModel}), and $\varepsilon\colon \mathcal{S}\to\mathcal{P}(\mathcal{S})$ be a function such that $s \in \varepsilon(s)$. We call $\varepsilon$ uncertainty specification. \emph{Uncertainty injection} is a construction that returns the Event-B model $\mathcal{M}^{\varepsilon}$ shown in Fig.~\ref{fig:uncertaintyInjectedModel}.
\end{mydefinition}

\begin{figure}[tbp]
    \centering
\begin{lstlisting}
Machine $\mathcal{M^{\varepsilon}}$
State space $\mathcal{S} \times \mathcal{S}$
Invariants ($I^\varepsilon \subseteq\mathcal{S} \times \mathcal{S} \; \mathrm{such\;that} \; \tuple{s,\hat{s}}\in I^\varepsilon \Longrightarrow \pi^{\varepsilon}(\tuple{s,\hat{s}})\subseteq I^\varepsilon$)
  Safety invariant $I^{\safety,\varepsilon}(\tuple{s,\hat{s}}) \teq I^{\safety}(s)$ /* events may violate this */ `\label{line:UncertaintyInjectedSafetyInvariant}`
  Uncertainty invariant $I^{\uncertainty,\varepsilon}(\tuple{s,\hat{s}}) \teq \bigl(s \in \varepsilon({\hat{s}})\bigr)$ `\label{line:UncertaintyInjectedUncertaintyInvariant}`
Initial states $A^{\varepsilon}_{0} \teq \{\tuple{s, \hat{s}}\, \big|\, s \in A_{0} \wedge s \in \varepsilon(\hat{s})\}$ `\label{line:UncertaintyInjectedUncertaintyInitState}`
Transition function $\pi^\varepsilon: \mathcal{S} \times \mathcal{S} \rightarrow \mathcal{P}(\mathcal{S} \times \mathcal{S})$, given by $\begin{array}[]{rl} \pi^\varepsilon(\tuple{s,\hat{s}})\teq    & \bigcup\bigl\{\,A^{\plant,\varepsilon}_{i}(\tuple{s,\hat{s}},p)\,\big|\,i\in[1,N_\plant], p\in P^{\plant,\varepsilon}_{i},  G^{\plant,\varepsilon}_{i}(\tuple{s,\hat{s}},p) \,\bigr\} \\ &\cup \bigcup\bigl\{\,A^{\controller,\varepsilon}_{i}(\tuple{s,\hat{s}},p)\,\big|\,i\in[1,N_\controller], p\in P^{\controller,\varepsilon}_{i}, G^{\controller,\varepsilon}_{i}(\tuple{s,\hat{s}},p)\,\bigr\}, \end{array}$ where
  Plant event $E^{\plant,\varepsilon}_{i} \quad (\text{where }i\in[1,N_\plant])$
    Parameter set $P^{\plant,\varepsilon}_{i} \teq P^{\plant}_{i}$
    Guard $G^{\plant,\varepsilon}_{i}(\tuple{s,\hat{s}},p) \tiff G^{\plant}_{i}(s,p)$ `\label{line:uncertaintyInjected:plantGuard}`
    Action $A^{\plant,\varepsilon}_{i}(\tuple{s,\hat{s}},p) \teq \{\tuple{s',\hat{s}'} \,\big|\, s' \in A^{\plant}_{i}(s,p) \wedge s' \in \varepsilon(\hat{s}')\}$ `\label{line:UncertaintyInjectedPlantAction}`
  Controller event $E^{\controller,\varepsilon}_{i}\quad (\text{where }i\in[1,N_\controller])$
    Parameter set $P^{\controller,\varepsilon}_{i} \teq P^{\controller}_{i}$
    Guard $G^{\controller,\varepsilon}_{i}(\tuple{s,\hat{s}},p) \tiff G^{\controller}_{i}(\hat{s},p)$ `\label{line:uncertaintyInjected:controllerGuard}`
    Action $A^{\controller,\varepsilon}_{i}(\tuple{s,\hat{s}},p) \teq \{\tuple{s',\hat{s}'} \,\big|\, s' \in A^{\controller}_{i}(s,p) \wedge s' \in \varepsilon(\hat{s}')\}$ `\label{line:UncertaintyInjectedCtrlAction}`
Subject to $\mathit{partitioning}$: $\forall \tuple{s,\hat{s}}\in\mathcal{S}\times\mathcal{S}.\,\mathop{\exists !} i\in [1,N_\controller].\,\exists p\in P^{\controller,\varepsilon}_{i}.\, G^{\controller,\varepsilon}_{i}(\tuple{s,\hat{s}}, p)$ `\label{line:UncertaintyInjectedPartitioning}`
 \end{lstlisting}
    \caption{The controller-plant model $\mathcal{M}^{\varepsilon}$ given by uncertainty injection from $\mathcal{M}$ (Fig.~\ref{fig:controllerPlantModel}) and  $\varepsilon\colon \mathcal{S}\to\mathcal{P}(\mathcal{S})$. 
Here $s,s'$ are \emph{true} states while $\hat{s},\hat{s}'$ are \emph{perceived} states. Note that $\mathcal{M}^{\varepsilon}$ may not preserve $I^{\safety,\varepsilon}$ due to the uncertainty.}
    \label{fig:uncertaintyInjectedModel}
\end{figure}

The key difference of $\mathcal{M}^{\varepsilon}$ from $\mathcal{M}$ (Fig.~\ref{fig:controllerPlantModel})
is that the state space $\mathcal{S}$ is duplicated--- i.e., $\mathcal{S}\times\mathcal{S}$ is the state space of $\mathcal{M}^{\varepsilon}$. In $\tuple{s,\hat{s}}\in \mathcal{S}\times\mathcal{S}$, $s$ is a \emph{true} state and $\hat{s}$ is a \emph{perceived} state.
The rest of the model $\mathcal{M}^{\varepsilon}$ closely follows $\mathcal{M}$, but whether a plant event $E^{\plant,\varepsilon}_{i}$ is enabled or not is decided based on the true state $s$ (Line~\ref{line:uncertaintyInjected:plantGuard}); while, the guard of a controller event $E^{\controller,\varepsilon}_{i}$  looks at the perceived state $\hat{s}$  (Line~\ref{line:uncertaintyInjected:controllerGuard}). Note, however, that, all actions $A^{\plant,\varepsilon}_{i}$ and $A^{\controller,\varepsilon}_{i}$ act on true states ($s$ and $s'$). In particular, controller actions are assumed to operate on the plant (i.e., the physical reality) via actuators. In Line~\ref{line:UncertaintyInjectedSafetyInvariant}, the safety invariant $I^{\safety,\varepsilon}$ checks if the \emph{true} state $s$ is safe.

The uncertainty specification $\varepsilon$ occurs in Lines~\ref{line:UncertaintyInjectedUncertaintyInvariant}, \ref{line:UncertaintyInjectedUncertaintyInitState}, \ref{line:UncertaintyInjectedPlantAction}, \ref{line:UncertaintyInjectedCtrlAction}. Lines~\ref{line:UncertaintyInjectedPlantAction} \&~\ref{line:UncertaintyInjectedCtrlAction} model the assumption that perception is made after each action with respect to the current true state ($s' \in \varepsilon(\hat{s}')$)---this means in particular that perception errors do not accumulate over time. The uncertainty invariant is added in $\mathcal{M}^{\varepsilon}$  (Line~\ref{line:UncertaintyInjectedUncertaintyInvariant}); this is maintained by the definition of actions (Lines~\ref{line:UncertaintyInjectedPlantAction} \&~\ref{line:UncertaintyInjectedCtrlAction}).  
We also note that the partitioning requirement (Line~\ref{line:UncertaintyInjectedPartitioning}) for $\mathcal{M}^{\varepsilon}$ remains satisfied.

Although the original model $\mathcal{M}$ is ``safe''
(in the Event-B sense of \emph{invariant preservation}, see~\S{}\ref{sec:ctrlPlantModel}),
the uncertainty-injected model $\mathcal{M}^{\varepsilon}$  may not be invariant-preserving. In \S{}\ref{sec:robustification}, we present syntactic model transformations to make it safe.

\begin{myexample}[The heater model $\mathcal{M}_{\mathrm{ht0}}^{\varepsilon_{0}}$]
 Fig.~\ref{fig:exampleInjected} is the model given by injecting the uncertainty to $\mathcal{M}_{\mathrm{ht0}}$ (Fig.~\ref{fig:exampleControllerPlant}). Here the uncertainty specification $\varepsilon_{0}$ is
 \begin{displaymath}
  \varepsilon_{0}\colon \mathcal{S}\rightarrow \mathcal{P}(\mathcal{S}),\; \tuple{\hsturn,\hstemp} \mapsto \{\tuple{\sturn,\stemp} \,\big|\, \sturn = \hsturn \wedge \stemp \in [\hstemp-3, \hstemp+3]\}.
 \end{displaymath}
 This specifies that sensed values of temperature can have errors up to 3$^\circ$C.

 The controller does not preserve the safety invariant $I^{\safety,\varepsilon_{0}}$ (Line~\ref{line:exampleInjected:safetyInvariant}).
 For example, when $\stemp = 32$ and $\hstemp = 29$, the event $\mathsf{ctrl\_heat}$ (Line~\ref{line:exampleInjected:event}) can fire with parameter $\pdh = 11$--- i.e., with perceived $\hstemp=29$ and maximum safe temperature $40$, the controller thinks that it can raise the temperature by $11$. 
 This leads $\stemp$ to $43$, violating the safety invariant $30 \leq \stemp \leq 40$.
\end{myexample}

\begin{figure}[tbp]
 \centering
\begin{lstlisting}
Machine $\mathcal{M}_{\mathrm{ht0}}^{\varepsilon_{0}}$
  $\ldots$
  Invariants
    Safety invariant $I^{\safety,\varepsilon_{0}} = \{\tuple{\tuple{\sturn,\stemp}, \tuple{\hsturn,\hstemp}} \,\big|\, \sturn = \controller \Longrightarrow 30 \leq \stemp \leq 40\}$`\label{line:exampleInjected:safetyInvariant}`
    Uncertainty invariant $I^{\uncertainty,\varepsilon_{0}} = \{\tuple{\tuple{\sturn,\stemp}, \tuple{\hsturn,\hstemp}} \,\big|\, \tuple{\sturn,\stemp} \in \varepsilon_{0}(\tuple{\hsturn,\hstemp})\}$
  $\ldots$
  Controller event $E_{1}^{\controller,\varepsilon_{0}}$                /* $\mathsf{ctrl\_heat}$ */`\label{line:exampleInjected:event}`
    Parameter set $P_{1}^{\controller,\varepsilon_{0}} = \mathbb{Z}$                /* parameter $\pdh$ */
    Guard $G_{1}^{\controller,\varepsilon_{0}}\tuple{\tuple{\tuple{\sturn,\stemp}, \tuple{\hsturn,\hstemp}},\pdh} \iff$ $\hstemp < 30 \wedge 30 \leq \hstemp + \pdh \leq 40$`\label{line:exampleInjected:guard}`
    Action $A_{1}^{\controller,\varepsilon_{0}}\tuple{\tuple{\tuple{\sturn,\stemp}, \tuple{\hsturn,\hstemp}},\pdh} = \{\tuple{\tuple{\sturn',\stemp'}, \tuple{\hsturn',\hstemp'}} \,\big|\,$ $\sturn' = \controller \wedge \stemp' = \stemp + \pdh \wedge \tuple{\sturn',\stemp'} \in \varepsilon_{0}(\tuple{\hsturn',\hstemp'})\}$
  $\ldots$
 \end{lstlisting}
 \caption{The heater model $\mathcal{M}_{\mathrm{ht0}}^{\varepsilon_{0}}$ produced by uncertainty injection}
 \label{fig:exampleInjected}
\end{figure}

\section{Robustification}
\label{sec:robustification}
We propose two syntactic transformations that modify the uncertainty-injected controller for the purpose of regaining safety. They are called \emph{action-preserving robustification} $(\underline{\phantom{x}})^{\mathsf{pR}}$ and \emph{action-repurposing robustification} $(\underline{\phantom{x}})^{\mathsf{rR}}$, respectively.

\subsection{Types of Robustified Events}
The basic idea is as follows, common to the two robustification transformations. 

\begin{figure}[tbp]
 \centering
 \includegraphics[width=.7\linewidth]{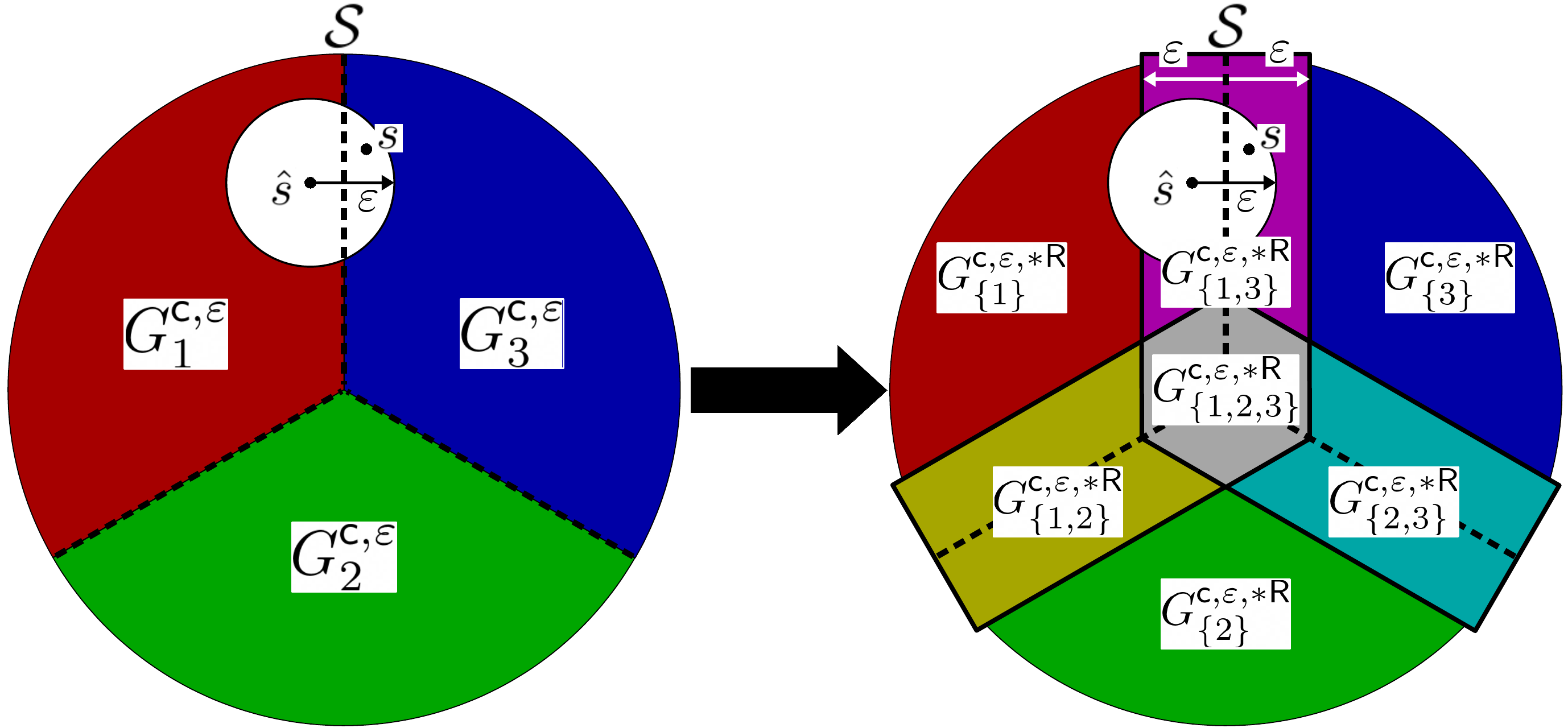}
 \caption{Uncertainty robustification}
 \label{fig:robustification}
\end{figure}

Assume the situation on the left in Fig.~\ref{fig:robustification}. If the controller was sure that the true state $s$ belonged to the region of $G^{\controller,\varepsilon}_{1}$,\footnote{This is the same as $G^{\controller}_{1}$, see Fig.~\ref{fig:uncertaintyInjectedModel}, Line~\ref{line:uncertaintyInjected:controllerGuard}.} then the controller could take the action $A^{\controller,\varepsilon}_{1}$. This way the controller can achieve the system's safety, since the original model $\mathcal{M}$ is safe. (Note that we implicitly rely on the \emph{partitioning} requirement of $\mathcal{M}$, Fig.~\ref{fig:controllerPlantModel}, Line~\ref{line:CPModelPartitioning}).
Unfortunately, the controller cannot be sure that the true state $s$ belongs to the region of $G^{\controller,\varepsilon}_{1}$ because, due to uncertainty, the set $\varepsilon(\hat{s})$ of potential true states overlaps with the region of another guard $G^{\controller,\varepsilon}_{3}$. Therefore, it is not clear from just looking at the perceived state $\hat{s}$ whether the controller should take the controller action $A^{\controller,\varepsilon}_{1}$ or  $A^{\controller,\varepsilon}_{3}$.  

To overcome the challenge, we first refine the partitioning of the state space---so that each compartment stands for the set of controller actions that are potentially enabled. For example, on the right in Fig.~\ref{fig:robustification}, we have seven compartments, such as $\{1,3\}$ for ``the true state $s$ must be  either in $G^{\controller,\varepsilon}_{1}$ or in $G^{\controller,\varepsilon}_{3}$.''

We create new events for the new compartments that arise this way (i.e., for those states that potentially enable multiple controller actions). These new events are called \emph{heterogeneous events}. For example, on the right in Fig.~\ref{fig:robustification}, we have four heterogeneous events 
$E^{\controller,\varepsilon}_{\{1,2\}},
E^{\controller,\varepsilon}_{\{1,3\}},
E^{\controller,\varepsilon}_{\{2,3\}},
E^{\controller,\varepsilon}_{\{1,2,3\}}$, in addition to the \emph{homogeneous events} $E^{\controller,\varepsilon}_{\{1\}}, E^{\controller,\varepsilon}_{\{2\}}, E^{\controller,\varepsilon}_{\{3\}}$, which are for cases where the controller is sure that it can use a particular action.

There are different ways that the actions of these heterogeneous events can be defined, leading to the two robustification methods presented in~\S{}\ref{sec:robustification-preserving}--\ref{sec:robustification-repurposing}. 
\begin{itemize}
 \item 
 In \emph{action-preserving} robustification $(\underline{\phantom{x}})^{\mathsf{pR}}$,
 the set of states reachable with a heterogeneous event $E^{\controller,\varepsilon,\mathsf{pR}}_{\{i_{1},\dotsc,i_{k}\}}$ is the \emph{intersection} of those reachable with $E^{\controller,\varepsilon}_{i_{1}},\dotsc, E^{\controller,\varepsilon}_{i_{k}}$.
 Therefore, the action of a heterogeneous event $E^{\controller,\varepsilon,\mathsf{pR}}_{\{i_{1},\dotsc,i_{k}\}}$ satisfies all requirements satisfied by
 $A^{\controller,\varepsilon}_{i_{1}},\dotsc,
 A^{\controller,\varepsilon}_{i_{k}}
 $.
 This way, the system generated with this method $\mathcal{M}^{\varepsilon,\mathsf{pR}}$ can be forward-simulated\footnote{This does not mean the refinement in Event-B, which requires every concrete event to have guards stronger than guards of abstract events.} by the original system $\mathcal{M}$, that is, any execution trace of $\mathcal{M}^{\varepsilon,\mathsf{pR}}$ is an execution trace of $\mathcal{M}$ ($s \in \epsilon(\hat{s}) \wedge \tuple{s', \hat{s}'} \in \pi^{\epsilon, \mathsf{pR}}(\tuple{s, \hat{s}})
\Longrightarrow s' \in \pi(s)$). In particular,  $\mathcal{M}^{\varepsilon,\mathsf{pR}}$ is safe since so is $\mathcal{M}$.
 \item
 In \emph{action-repurposing} robustification $(\underline{\phantom{x}})^{\mathsf{rR}}$,
 the action of a heterogenerous event $E^{\controller,\varepsilon,\mathsf{rR}}_{\{i_{1},\dotsc,i_{k}\}}$ does not consider satisfying all requirements satisfied by $A^{\controller,\varepsilon}_{i_{1}},\dotsc, A^{\controller,\varepsilon}_{i_{k}}$.
 The event $E^{\controller,\varepsilon,\mathsf{rR}}_{\{i_{1},\dotsc,i_{k}\}}$ uses at least \emph{one} (but not necessarily all) of $A^{\controller,\varepsilon}_{i_{1}},\dotsc, A^{\controller,\varepsilon}_{i_{k}}$ with parameters that are \emph{guaranteed to preserve the safety invariant} $I^{\safety,\varepsilon}$ regardless of which guard can be satisfied by the true state.
 Therefore, an action originally from  $E^{\controller,\varepsilon}_{i}$ may be invoked from the region of the guard of $E^{\controller,\varepsilon}_{j}$ with $i\neq j$, making the behavior of the resulting model different from that of $\mathcal{M}$. In this way, actions from one event may be ``repurposed'' for another event, hence the name of the method.
\end{itemize}

In our workflow (Fig.~\ref{fig:methods-overview}), we prefer the action-preserving robustification since it yields a controller that can be forward-simulated by the original one. In case it is not feasible (i.e., when no action is shared by the events  $E^{\controller,\varepsilon}_{i_{1}},\dotsc,
 E^{\controller,\varepsilon}_{i_{k}}
 $), we try the action-repurposing robustification.

\subsection{Action-Preserving Robustification}
\label{sec:robustification-preserving}
\begin{mydefinition}[Action-preserving robustification $(\underline{\phantom{x}})^{\mathsf{pR}}$]
\label{def:actionPresRobustif}
\emph{Action-preserving robustification} is a construction that 
 takes an Event-B model $\mathcal{M}^{\varepsilon}$ as shown in Fig.~\ref{fig:uncertaintyInjectedModel} as input, and
returns  the Event-B model $\mathcal{M}^{\varepsilon,\mathsf{pR}}$ in Fig.~\ref{fig:uncertaintyPreservingRobustifiedModel}.
In Fig.~\ref{fig:uncertaintyPreservingRobustifiedModel} (and elsewhere below), we use the following functions.
\begin{itemize}
 \item The function $\eIdx\colon \mathcal{S}\to [1,N_\controller]$ returns, for each state $s\in \mathcal{S}$ (in the original system $\mathcal{M}$), the index of the unique controller event enabled at $s$ in the original model $\mathcal{M}$. That is, $\exists p\in P^{\controller}_{\eIdx(s)}.\, G^{\controller}_{\eIdx(s)}(s,p)$ holds. 
 \item The function $\ePar$  takes a state $s\in \mathcal{S}$ and returns the set of parameter values that are compatible, that is, $\ePar(s)=\{p\in P^{\controller}_{\eIdx(s)}\mid G^{\controller}_{\eIdx(s)}(s,p)\}$.
 \item The function $\eParEpsi$ takes a state $\hat{s}\in \mathcal{S}$  and returns
$\eParEpsi(\hat{s}) = \bigcap \{\ePar(\tilde{s})\mid \eIdx(\tilde{s})=i, \tilde{s} \in \varepsilon(\hat{s})\}$, i.e., the set of parameter values compatible with any state $\tilde{s}$ that is in the $\varepsilon$-neighborhood of $\hat{s}$ and enables $E^{\controller}_{i}$. 
\end{itemize}

\end{mydefinition}

\begin{figure}[tbp]
 \centering
\begin{lstlisting}
Machine $\mathcal{M}^{\varepsilon,\mathsf{pR}}$
(State space, invariant, and initial states are the same as $\mathcal{M}^{\varepsilon}$)
  Transition function $\pi^{\varepsilon,\mathsf{pR}}: \mathcal{S} \times \mathcal{S} \rightarrow \mathcal{P}(\mathcal{S} \times \mathcal{S})$, given by
    $\begin{array}[]{l} 
 {\pi^{\varepsilon,\mathsf{pR}}}(\tuple{s,\hat{s}})= \\
 \quad
  \bigcup
\bigl\{\,A^{\plant,\varepsilon}_{i}(\tuple{s,\hat{s}},p^{\plant})\,\big|\,i\in[1,N_\plant], 
p^{\plant}\in P^{\plant,\varepsilon}_{i},  G^{\plant,\varepsilon}_{i}(\tuple{s,\hat{s}},p^{\plant}) \,\bigr\}
     \\\quad\cup \bigcup\bigl\{\,A_{u}^{\controller,\varepsilon,\mathsf{pR}}(\tuple{s,\hat{s}},p^{\controller})\,\big|\,u \in \mathcal{P}([1,N_\controller]) \setminus \emptyset, 
 p^{\controller}\in P_{u}^{\controller,\varepsilon,\mathsf{pR}}, G_{u}^{\controller,\varepsilon,\mathsf{pR}}(\tuple{s,\hat{s}},p^{\controller})\,\bigr\}, \end{array}$
    Plant event $E_{i}^{\plant,\varepsilon,\mathsf{pR}} \teq E_{i}^{\plant,\varepsilon}$
    Controller event $E_{u}^{\controller,\varepsilon,\mathsf{pR}} \quad(\text{where } u \in \mathcal{P}([1,N_\controller]) \setminus \emptyset)$`\label{line:PreservingRobustificationEvent}`
      Parameter set $P_{u}^{\controller,\varepsilon,\mathsf{pR}} \teq \prod_{i \in u} (P^{\controller}_{i} \cup \{\bot_{i}\})$`\label{line:PreservingRobustificationParameterDef}`
      Guard $G_{u}^{\controller,\varepsilon,\mathsf{pR}}(\tuple{s,\hat{s}},p^{\controller}) \tiff$
        $u=\{\eIdx(\tilde{s})\mid \tilde{s}\in \varepsilon(\hat{s})\}$`\label{line:PreservingRobustificationUGuard}` /* $\tilde{s}$: a potential true state */
        $\wedge$ $\forall i\in u.\, ((\eParEpsi(\hat{s})\neq \emptyset \Longrightarrow p^{\controller}_{i}\in \eParEpsi(\hat{s})) \wedge (\eParEpsi(\hat{s})= \emptyset \Longrightarrow p^{\controller}_{i}=\bot_{i}))$`\label{line:PreservingRobustificationParameter}`
        $\wedge$ $\forall \tilde{s} \in \varepsilon(\hat{s}) .\, (\bigcap_{i\in u} A^{\controller}_{i}(\tilde{s},p^{\controller}_{i}) \neq \emptyset)$`\label{line:PreservingRobustificationNonemptyness}`
      Action ${A_{u}^{\controller,\varepsilon,\mathsf{pR}}}:(\mathcal{S} \times \mathcal{S}) \times P_{u}^{\controller,\varepsilon,\mathsf{pR}} \rightarrow \mathcal{P}(\mathcal{S} \times \mathcal{S})$
        $A_{u}^{\controller,\varepsilon,\mathsf{pR}}(\tuple{s,\hat{s}}, p^{\controller}) = \{\tuple{s',\hat{s}'}\mid s' \in \bigcap_{i\in u} A^{\controller}_{i}(s,p^{\controller}_{i}) \wedge s' \in \varepsilon(\hat{s}')\}$`\label{line:PreservingRobustificationAction}`
  Subject to $\mathit{partitioning}$: $\forall \tuple{s,\hat{s}}\in\mathcal{S}\times\mathcal{S}.\,\mathop{\exists !} u\in \mathcal{P}([1,N_\controller]).\,\exists p^{\controller} \in P_{u}^{\controller,\varepsilon,\mathsf{pR}}.\, G_{u}^{\controller,\varepsilon,\mathsf{pR}}(\tuple{s,\hat{s}}, p^{\controller})$ `\label{line:PreservingRobustificationPartitioning}`
 \end{lstlisting}
 \caption{A controller-plant model $\mathcal{M}^{\varepsilon,\mathsf{pR}}$ produced by action-preserving robustification from $\mathcal{M}^{\varepsilon}$ from Fig.~\ref{fig:uncertaintyInjectedModel}}
 \label{fig:uncertaintyPreservingRobustifiedModel}
\end{figure}

The parameter value for the index $i \in u$ may be $\bot_i$ (Line~\ref{line:PreservingRobustificationParameterDef}) such that $\forall s \in \mathcal{S}. A^{\controller}_{i}(s,\bot_i) = \emptyset$. $P^{\controller}_{i} = \bot_i$ means that there is no $i$-th parameter that satisfies constraints on parameters for safety and feasibility (Lines~\ref{line:PreservingRobustificationParameter}--\ref{line:PreservingRobustificationNonemptyness}).

\begin{theorem}
\label{thm:actionPreservingFeasibility}
Regarding the model $\mathcal{M}^{\varepsilon,\mathsf{pR}}$ in Def.~\ref{def:actionPresRobustif} (Fig.~\ref{fig:uncertaintyPreservingRobustifiedModel}), assume the following condition (i.e., for all events, there exist parameter values such that they are compatible with all possible states under the $\varepsilon$-uncertainty and there exist actions common in all original actions) is satisfied.
 \begin{align*}\begin{array}{l}
  		 \forall u \in \mathcal{P}([1,N_\controller]) .\, \forall \tuple{s, \hat{s}} \in \mathcal{S} \times \mathcal{S} .\, (s \in \varepsilon(\hat{s}) \wedge u = \{\eIdx(\tilde{s}) | \tilde{s} \in \varepsilon(\hat{s})\} \Longrightarrow
 		\\
  		 \quad \exists p^{\controller} = \tuple{p^{\controller}_{i_1}, \ldots, p^{\controller}_{i_k}} \in P_{u}^{\controller,\varepsilon,\mathsf{pR}} .\, 
       \\
 \quad \quad
  ((\forall \tilde{s} \in \varepsilon(\hat{s}) .\, p^{\controller}_{\eIdx(\tilde{s})} \in \ePar(\tilde{s})) 
  \wedge  (\exists s' \in \mathcal{S} .\, \forall i \in u .\, s' \in A^{\controller}_{i}(s,p^{\controller}_{i})))).
\end{array}
 \end{align*}
Then  $\mathcal{M}^{\varepsilon,\mathsf{pR}}$ satisfies the partitioning requirement (Fig.~\ref{fig:uncertaintyPreservingRobustifiedModel}, Line~\ref{line:PreservingRobustificationPartitioning}), and is invariant-preserving and  feasible (Def.~\ref{def:invariantPreservationFeasibility}). 
\end{theorem}

We judge the success of the action-preserving robustification by the condition in Thm.~\ref{thm:actionPreservingFeasibility}. If it fails, then we try the action-repurposing robustification (Fig.~\ref{fig:methods-overview}).

\begin{myexample}[The heater model $\mathcal{M}_{\mathrm{ht0}}^{\varepsilon_{0},\mathsf{pR}}$]
\label{ex:robustifiedPre}
 Fig.~\ref{fig:exampleRobustifiedPre} is an excerpt from the model obtained by applying the action-preserving robustification to $\mathcal{M}_{\mathrm{ht0}}^{\varepsilon_{0}}$ (Fig.~\ref{fig:exampleInjected}) showing the heterogeneous event $E_{\{1,2\}}^{\varepsilon,\controller,\mathsf{pR}}$ generated from the event $\mathsf{ctrl\_heat}$ and the event $\mathsf{ctrl\_keep\_safe}$ from $\mathcal{M}_{\mathrm{ht0}}$.
 Constraints on the perceived temperature (Lines~\ref{line:examplePreserving:stateGuardBegin}--\ref{line:examplePreserving:stateGuardEnd}) mean that $u = \{1,2\}$ in this event (Line~\ref{line:PreservingRobustificationUGuard} of Fig.\ref{fig:uncertaintyPreservingRobustifiedModel}).
 Constraints on parameters (Lines~\ref{line:examplePreserving:paramGuardBegin}--\ref{line:examplePreserving:paramGuardEnd}) mean that $\pdh$ and $\pdt$ are compatible with every state $\tstemp$ around $\hstemp$ (Line~\ref{line:PreservingRobustificationParameter} of Fig.\ref{fig:uncertaintyPreservingRobustifiedModel}).
 Line~\ref{line:examplePreserving:actionNonemptyGuard} means that there are common actions in $\mathsf{ctrl\_heat}$ and $\mathsf{ctrl\_keep\_safe}$ (Line~\ref{line:PreservingRobustificationNonemptyness} of Fig.\ref{fig:uncertaintyPreservingRobustifiedModel}).

 The event $E_{\{1,2\}}^{\varepsilon,\controller,\mathsf{pR}}$ is indeed feasible and it preserves the safety invariant $I^{\safety}$ as all other events of $\mathcal{M}_{\mathrm{ht0}}^{\varepsilon_{0},\mathsf{pR}}$ do.
 Lines~\ref{line:examplePreserving:stateGuardBegin}--\ref{line:examplePreserving:stateGuardEnd} mean that $27 \leq \hstemp < 33$.
 Also, Lines~\ref{line:examplePreserving:paramGuardBegin}--\ref{line:examplePreserving:paramGuardEnd} mean that $33-\hstemp \leq \pdh \leq 10$ and $0 \leq \pdt \leq 37-\hstemp$.
 In addition, line~\ref{line:examplePreserving:actionNonemptyGuard} requests that $\pdh = \pdt$, thus $33 - \hstemp \leq \pdh = \pdt \leq 37 - \hstemp$.
 Since $\stemp \in [\hstemp-3, \hstemp+3]$, we can guarantee that $I^{\safety}$ is preserved, namely $30 \leq \stemp + \pdh = \stemp + \pdt \leq 40$.
 For example, if $\hstemp = 29$, then $26 \leq \tstemp \leq 32$.
 In case of $26 \leq \tstemp < 30$, the event $\mathsf{ctrl\_heat}$ would heat to increase the temperature by $\pdh$ where $4 \leq \pdh \leq 10$ (Line \ref{line:examplePreserving:paramGuardBegin}).
 Otherwise (i.e., $30 \leq \tstemp \leq 32$), the event $\mathsf{ctrl\_keep\_safe}$ would change the temperature for $\pdt$ where $0 \leq \pdt \leq 8$ (Line \ref{line:examplePreserving:paramGuardEnd}).
 The common actions here are changing temperature by $\pdh = \pdt \in [4,8]$, which is safe for all $\tstemp \in [26,32]$.
\end{myexample}

\begin{figure}[tbp]
 \centering
\begin{lstlisting}
Machine $\mathcal{M}_{\mathrm{ht0}}^{\varepsilon_{0},\mathsf{pR}}$
  $\ldots$
  Controller event $E_{\{1,2\}}^{\controller,\varepsilon_{0},\mathsf{pR}}$          /* $\mathsf{ctrl\_heat\_keep\_safe\_hetero}$ */
    Parameter set $P_{\{1,2\}}^{\controller,\varepsilon_{0},\mathsf{pR}} = \mathbb{Z} \times \mathbb{Z}$                /* parameter $\pdh$ and $\pdt$ */
    Guard $G_{\{1,2\}}^{\controller,\varepsilon_{0},\mathsf{pR}} \tuple{\tuple{\tuple{\sturn,\stemp}, \tuple{\hsturn,\hstemp}},\tuple{\pdh,\pdt}} \iff$
      $\; \; \; \forall \tstemp \in [\hstemp-3, \hstemp+3] .\, (\tstemp < 30 \vee 30 \leq \tstemp \leq 40)$`\label{line:examplePreserving:stateGuardBegin}`
      $\wedge \; \exists \tstemp \in [\hstemp-3, \hstemp+3] .\, (\tstemp < 30)$
      $\wedge \; \exists \tstemp \in [\hstemp-3, \hstemp+3] .\, (30 \leq \tstemp \leq 40)$`\label{line:examplePreserving:stateGuardEnd}`
      $\wedge \; \forall \tstemp \in [\hstemp-3, \hstemp+3] .\, (\tstemp < 30 \Longrightarrow 30 \leq \tstemp + \pdh \leq 40)$`\label{line:examplePreserving:paramGuardBegin}`
      $\wedge \; \forall \tstemp \in [\hstemp-3, \hstemp+3] .\, (30 \leq \tstemp \leq 40 \Longrightarrow 30 \leq \tstemp + \pdt \leq 40)$`\label{line:examplePreserving:paramGuardEnd}`
      $\wedge$ $\forall \tstemp \in [\hstemp-3, \hstemp+3] .\,$ $(\{\tstemp' \big| \tstemp' = \tstemp + \pdh\} \cap \{\tstemp' \big| \tstemp' = \tstemp + \pdt\}) \neq \emptyset$`\label{line:examplePreserving:actionNonemptyGuard}`
    Action $A_{\{1,2\}}^{\controller,\varepsilon_{0},\mathsf{pR}} \tuple{\tuple{\tuple{\sturn,\stemp}, \tuple{\hsturn,\hstemp}},\tuple{\pdh,\pdt}} = \{\tuple{\tuple{\sturn',\stemp'}, \tuple{\hsturn',\hstemp'}} \,\big|\,$ $\sturn' = \controller \wedge \stemp' = \stemp + \pdh \wedge \stemp' = \stemp + \pdt \wedge \tuple{\sturn',\stemp'} \in \varepsilon_{0}(\tuple{\hsturn',\hstemp'})\}$
  $\ldots$
 \end{lstlisting}
 \caption{The heater model $\mathcal{M}_{\mathrm{ht0}}^{\varepsilon_{0},\mathsf{pR}}$ produced by action-preserving robustification}
 \label{fig:exampleRobustifiedPre}
\end{figure}

\subsection{Action-Repurposing Robustification}
\label{sec:robustification-repurposing}

\begin{mydefinition}[Action-repurposing robustification $(\underline{\phantom{x}})^{\mathsf{rR}}$]
\label{def:actionRepurRobustif}
\emph{Action-repurposing robustification} is a construction that
takes an Event-B model $\mathcal{M}^{\varepsilon}$ as shown in Fig.~\ref{fig:uncertaintyInjectedModel} as input, and
returns the Event-B model $\mathcal{M}^{\varepsilon,\mathsf{rR}}$ (Fig.~\ref{fig:actionRepurposingRobustifiedModel}).
In Fig.~\ref{fig:actionRepurposingRobustifiedModel} (and elsewhere below), we use the following function.
\begin{itemize}
 \item The function $\eSafePari$ takes a state $\hat{s}\in \mathcal{S}$ and a safety invariant $I$ and returns
       $\bigcap_{\tilde{s} \in \varepsilon(\hat{s})} \{p | \emptyset \subset A^{\controller}_{i}(\tilde{s}, p) \subseteq I\}$,
       i.e., the set of parameter values that preserve the safety invariant $I$ when used with the action of the $i$-th event of $\mathcal{M}$ ($A^{\controller}_{i}$) at any state $\tilde{s}$ that is in the $\varepsilon$-neighborhood of $\hat{s}$ and enables $E^{\controller}_{i}$. 
\end{itemize}
\end{mydefinition}

\begin{figure}[tbp]
 \centering
\begin{lstlisting}
Machine $\mathcal{M}^{\varepsilon,\mathsf{rR}}$
(State space, invariant, and initial states are same as $\mathcal{M}^{\varepsilon}$)
Transition function $\pi^{\varepsilon,\mathsf{rR}}: \mathcal{S} \times \mathcal{S} \rightarrow \mathcal{P}(\mathcal{S} \times \mathcal{S})$, given by
  $\begin{array}[]{l} {\pi^{\varepsilon,\mathsf{rR}}}(\tuple{s,\hat{s}})= \\ \quad \bigcup \bigl\{\,A^{\plant,\varepsilon}_{i}(\tuple{s,\hat{s}},p^{\plant})\,\big|\,i\in[1,N_\plant], p^{\plant}\in P^{\plant,\varepsilon}_{i},  G^{\plant,\varepsilon}_{i}(\tuple{s,\hat{s}},p^{\plant}) \,\bigr\} \\\quad\cup \bigcup\bigl\{\,A_{u}^{\controller,\varepsilon,\mathsf{rR}}(\tuple{s,\hat{s}},p^{\controller})\,\big|\,u \in \mathcal{P}([1,N_\controller]) \setminus \emptyset, p^{\controller}\in P_{u}^{\controller,\varepsilon,\mathsf{rR}}, G_{u}^{\controller,\varepsilon,\mathsf{rR}}(\tuple{s,\hat{s}},p^{\controller})\,\bigr\}, \end{array}$
  Plant event $E_{i}^{\plant,\varepsilon,\mathsf{rR}} \teq E_{i}^{\plant,\varepsilon}$
  Controller event $E_{u}^{\controller,\varepsilon,\mathsf{rR}} \quad(\text{where } u \in \mathcal{P}([1,N_\controller]) \setminus \emptyset)$`\label{line:RepurposingRobustificationEvent}`
    Parameter set $P_{u}^{\controller,\varepsilon,\mathsf{rR}} \teq \prod_{i \in u} (P^{\controller}_{i} \cup \{\bot_{i}\})$`\label{line:RepurposingRobustificationParameterDef}`
    Guard $G_{u}^{\controller,\varepsilon,\mathsf{rR}}(\tuple{s,\hat{s}},p^{\controller}) \tiff$
      $u=\{\eIdx(\tilde{s})\mid \tilde{s}\in \varepsilon(\hat{s})\}$`\label{line:RepurposingRobustificationUGuard}` /* $\tilde{s}$: a potential true state */
      $\wedge$ $\forall i\in u.\, ((\eSafePari(\hat{s},I^{\safety})\neq \emptyset \Rightarrow p^{\controller}_{i}\in \eSafePari(\hat{s},I^{\safety})) \wedge (\eSafePari(\hat{s},I^{\safety})= \emptyset \Rightarrow p^{\controller}_{i}=\bot_{i}))$`\label{line:RepurposingRobustificationParamGuardBegin}`
      $\wedge$ $\exists i \in u .\, p^{\controller}_{i} \neq \bot_{i}$`\label{line:RepurposingRobustificationFeasibility}`
    Action ${A_{u}^{\controller,\varepsilon,\mathsf{rR}}}:(\mathcal{S} \times \mathcal{S}) \times P_{u}^{\controller,\varepsilon,\mathsf{rR}} \rightarrow \mathcal{P}(\mathcal{S} \times \mathcal{S})$
      $A_{u}^{\controller,\varepsilon,\mathsf{rR}}(\tuple{s,\hat{s}}, p^{\controller}) = \{\tuple{s',\hat{s}'} \big| i \in u \wedge p^{\controller}_{i} \neq \bot_{i} \wedge s' \in A^{\controller}_{i}(s, p^{\controller}_{i}) \wedge s' \in \varepsilon(\hat{s}')\}$`\label{line:RepurposingRobustificationAction}`
Subject to $\mathit{partitioning}$: $\forall \tuple{s,\hat{s}}\in\mathcal{S}\times\mathcal{S}.\,\mathop{\exists !} u\in \mathcal{P}([1,N_\controller]).\,\exists p^{\controller} \in P_{u}^{\controller,\varepsilon,\mathsf{rR}}.\, G_{u}^{\controller,\varepsilon,\mathsf{rR}}(\tuple{s,\hat{s}}, p^{\controller})$`\label{line:RepurposingRobustificationPartitioning}`
 \end{lstlisting}
 \caption{A controller-plant model $\mathcal{M}^{\varepsilon,\mathsf{rR}}$ produced by action-repurposing robustification from $\mathcal{M}_{\varepsilon}$ from Fig.~\ref{fig:uncertaintyInjectedModel}}
 \label{fig:actionRepurposingRobustifiedModel}
\end{figure}

The model $\mathcal{M}^{\varepsilon,\mathsf{rR}}$ is the same as $\mathcal{M}^{\varepsilon,\mathsf{pR}}$ (Fig.~\ref{fig:uncertaintyPreservingRobustifiedModel}) except lines \ref{line:RepurposingRobustificationParamGuardBegin}, \ref{line:RepurposingRobustificationFeasibility}, and \ref{line:RepurposingRobustificationAction}.
For each original controller event $E^{\controller}_{i}$, the parameter of the event is restricted so that it satisfies the safety invariant $I^{\safety}$ for all possible states under the $\varepsilon$-uncertainty (Line~\ref{line:RepurposingRobustificationParamGuardBegin}).
The robustified controller uses \emph{one of} the events that have such parameter values (Line~\ref{line:RepurposingRobustificationAction}).
This guarantees that the safety invariant $I^{\safety}$ is satisfied for every possible true state (i.e., those in $\varepsilon(\hat{s})$).

\begin{theorem}
 Regarding the model $\mathcal{M}^{\varepsilon,\mathsf{rR}}$ in Def.~\ref{def:actionRepurRobustif}, assume the following condition (i.e., there exist original controller events and their parameter values that satisfy the safety at all possible states under the $\varepsilon$-uncertainty) is satisfied.
 \begin{align*}
  \begin{array}{l}
   \forall u \in \mathcal{P}([1,N_\controller]) .\, \forall \tuple{s, \hat{s}} \in \mathcal{S} \times \mathcal{S} .\, (s \in \varepsilon(\hat{s}) \wedge \{\eIdx(\tilde{s}) | \tilde{s} \in \varepsilon(\hat{s})\} \Longrightarrow \\
   \quad \exists i \in u, p^{\controller}_{i} \in P^{\controller}_{i} .\, (\forall \tilde{s} \in \varepsilon(\hat{s}) .\, A^{\controller}_{i}(\tilde{s},p^{\controller}_{i}) \subseteq I^{\safety})).
  \end{array}  
 \end{align*}
 Then $\mathcal{M}^{\varepsilon,\mathsf{rR}}$ satisfies the partition requirement (Fig.~\ref{fig:actionRepurposingRobustifiedModel}, line~\ref{line:RepurposingRobustificationPartitioning}), and is invariant-preserving and  feasible (Def.~\ref{def:invariantPreservationFeasibility}).
\end{theorem}

\begin{myexample}[The heater model $\mathcal{M}_{\mathrm{ht0}}^{\varepsilon_{0},\mathsf{rR}}$]
\label{ex:robustifiedRep}
 Fig.~\ref{fig:exampleRobustifiedRep} is an excerpt from the model obtained by applying the action-repurposing robustification to $\mathcal{M}_{\mathrm{ht0}}^{\varepsilon_{0}}$ (Fig.~\ref{fig:exampleInjected}), showing the heterogeneous event $E_{\{1,2\}}^{\varepsilon,\controller,\mathsf{rR}}$ constructed from $\mathsf{ctrl\_heat}$ and $\mathsf{ctrl\_keep\_safe}$.
The action of $\mathsf{ctrl\_heat}$ is adopted as the action of the event $E_{\{1,2\}}^{\controller,\varepsilon_{0},\mathsf{rR}}$ (Line~\ref{ex:robustifiedRepAction}).
The parameter $\pdh$ is restricted so that the safety invariant is preserved by the event even under the uncertainty (Line~\ref{ex:robustifiedRepParamConstraint}).
Thus, this event safely deals with the case where the controller is unsure if $\stemp < 30$ or $30 \leq \stemp \leq 40$ by \emph{repurposing} the action for the $\stemp < 30$ case.
\end{myexample}

\begin{figure}[tbp]
 \centering
\begin{lstlisting}
Machine $\mathcal{M}_{\mathrm{ht0}}^{\varepsilon_{0},\mathsf{rR}}$
  $\ldots$
  Controller event $E_{\{1,2\}}^{\controller,\varepsilon_{0},\mathsf{rR}}$          /* $\mathsf{ctrl\_heat\_keep\_safe\_hetero}$ */
    Parameter set $P_{\{1,2\}}^{\controller,\varepsilon_{0},\mathsf{rR}} = \mathbb{Z} \times \mathbb{Z}$                /* parameter $\pdh$ and $\pdt$ */
    Guard $G_{\{1,2\}}^{\controller,\varepsilon_{0},\mathsf{rR}} \tuple{\tuple{\tuple{\sturn,\stemp}, \tuple{\hsturn,\hstemp}},\tuple{\pdh,\pdt}} \iff$
      $\;\;\; \forall \tstemp \in [\hstemp-3, \hstemp+3] .\, (\tstemp < 30 \vee 30 \leq \tstemp \leq 40)$
      $\wedge \; \exists \tstemp \in [\hstemp-3, \hstemp+3] .\, \tstemp < 30$
      $\wedge \; \exists \tstemp \in [\hstemp-3, \hstemp+3] .\, 30 \leq \tstemp \leq 40$
      $\wedge \; \forall \tstemp \in [\hstemp-3, \hstemp+3] .\, 30 \leq \tstemp + \pdh \leq 40$`\label{ex:robustifiedRepParamConstraint}`
    Action $A_{\{1,2\}}^{\controller,\varepsilon_{0},\mathsf{rR}} \tuple{\tuple{\tuple{\sturn,\stemp}, \tuple{\hsturn,\hstemp}},\tuple{\pdh,\pdt}} = \{\tuple{\tuple{\sturn',\stemp'}, \tuple{\hsturn',\hstemp'}} \,\big|\,$ $\sturn' = \controller \wedge \stemp' = \stemp + \pdh \wedge \tuple{\sturn',\stemp'} \in \varepsilon_{0}(\tuple{\hsturn',\hstemp'})\}$`\label{ex:robustifiedRepAction}`
  $\ldots$
 \end{lstlisting}
 \caption{A heater model $\mathcal{M}_{\mathrm{ht0}}^{\varepsilon_{0},\mathsf{rR}}$ produced by action-repurposing robustification}
 \label{fig:exampleRobustifiedRep}
\end{figure}

\subsection{Checking Vacuity of Heterogeneous Events}
\label{sec:vacuousness}
A controller event of a robustified model corresponds to a non-empty subset of original controller events.
Therefore, if the original model has $n$ controller events, then the robustified model can have $2^n-1$ controller events.

However, there may be heterogeneous events of \emph{vacuous} cases.
For instance, in the robustified heater model (Fig.~\ref{fig:exampleRobustifiedPre}), the heterogeneous event $E^{\controller,\varepsilon_{0},\mathsf{pR}}_{\{1,2,3\}}$ (an event for when the controller is not sure if it should heat, keep safe, or cool) is vacuous because $\hstemp$ should satisfy $(\exists \tau \in \varepsilon_{\tau}(\widehat{temp}) .\, \tau < 30) \wedge (\exists \tau \in \varepsilon_{\tau}(\widehat{temp}) .\, 30 \leq \tau \leq 40) \wedge (\exists \tau \in \varepsilon_{\tau}(\widehat{temp}) .\, 40 < \tau
)$, where $\varepsilon_{\tau} = (\lambda \tau .\, [\tau-3, \tau+3])$; but, this is not satisfiable.
The vacuity of heterogeneous events depends on the uncertainty; for example, $E^{\controller,\varepsilon',\mathsf{pR}}_{\{1,2,3\}}$ is not vacuous when $\varepsilon'$ defines errors up to 7 because if $\hstemp = 35$ then $\stemp$ can be in the range $[35-7, 35+7]$.

Detecting and removing vacuous heterogeneous events is important because developers want meaningful descriptive models for reasoning.
In addition, it improves reasoning efficiency because it reduces the number of events.

\section{Implementation}
\label{sec:implementation}
\begin{figure}[tbp]
  \centering
  \includegraphics[width=.8\linewidth]{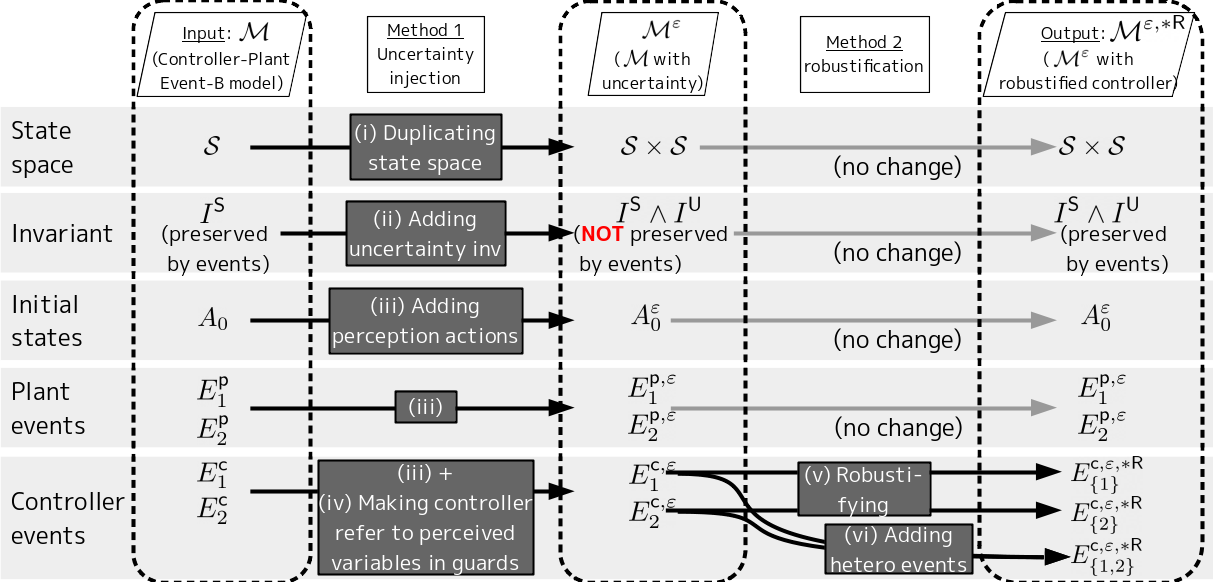}
  \caption{Overview of manipulations performed by our tool}
  \label{fig:manipulations_overview}
\end{figure}
Fig.~\ref{fig:manipulations_overview} is an overview of the model transformations used in the workflow of Fig.~\ref{fig:methods-overview}. Note that each transformation step is syntactic and thus can be automated. 
We implemented the workflow \footnote{Available at \url{http://research.nii.ac.jp/robustifier/}} as a plug-in of the Rodin Platform \cite{Abrial2010Rodin,EventBOrg}, which is the modeling environment of Event-B.
In the robustification process, it calculates (assisted by the Z3 SMT solver \cite{DeMouraZ3}) if robustification methods can generate invariant-preserving and feasible models. It also checks if each generated heterogeneous event is vacuous and thus should be removed (\S~\ref{sec:vacuousness}).

\begin{figure}[tb]

 \centering
\begin{lstlisting}
Machine $\mathcal{M}_{\mathrm{ht1}}$
  $\ldots$
  Controller event $E^{\controller}_{2}$                /* $\mathsf{ctrl\_keep\_safe\_eco}$ */
    Parameter set $P^{\controller}_{2} = \mathbb{Z}$                /* parameter $\pdt$ */
    Guard $G^{\controller}_{2}\tuple{\tuple{\sturn,\stemp},\pdt} \iff 30 \leq \stemp \leq 40$`\label{line:ex:ht1:compartment}`
      $\wedge$  $30 \leq \stemp+\pdt \leq 40$ $\wedge$ $-4 \leq \pdt \leq 4$ /* Only small changes */ `\label{line:ex:ht1:param}`
    Action $A^{\controller}_{2}\tuple{\tuple{\sturn,\stemp},\pdt} = \{\tuple{\sturn',\stemp'} \,\big|\, \sturn' = \controller \wedge \stemp' = \stemp+\pdt\}$
  $\ldots$
 \end{lstlisting}
 \caption{Model $\mathcal{M}_{\mathrm{ht1}}$: A variant of $\mathcal{M}_{\mathrm{ht0}}$ (Fig.~\ref{fig:exampleControllerPlant})}
 \label{fig:exampleVariant}

\end{figure}

\section{Case Study}
\label{sec:casestudies}

We demonstrate that our workflow helps developers to explore multiple levels of perceptual uncertainty.
Specifically, we give a \emph{parameterized} uncertainty specification and a model to the workflow and calculate the maximum level of uncertainty that generated controllers can tolerate.

Assume that we have a heater controller model $\mathcal{M}_{\mathrm{ht1}}$ (Fig.~\ref{fig:exampleVariant}).
$\mathcal{M}_{\mathrm{ht1}}$ is the same as $\mathcal{M}_{\mathrm{ht0}}$ (Fig.~\ref{fig:exampleControllerPlant}) except it has an \emph{ecological} ``keep\_safe'' functionality---
the change of the temperature $\pdt$ is limited to $[-4, +4]$ (Line~\ref{line:ex:ht1:param}).

When we choose a sensor module for $\mathcal{M}_{\mathrm{ht1}}$ from a series of modules with various prices and uncertainty (from cheap and more uncertain to expensive and less uncertain), the following question arises: \emph{How uncertain can the sensor module of $\mathcal{M}_{\mathrm{ht1}}$ be and still be safe?}
We show how we can answer this question with a manual analysis assisted by our automated workflow.
Here we assume that the series of sensor modules have parameterized uncertainty $\varepsilon_{\Delta t} = \lambda \stemp .\, [\stemp-\Delta t, \stemp+\Delta t]$, where $0 \leq \Delta t$.

\paragraph{Action-preserving robustification.}

The action-preserving robustification generates the model $\mathcal{M}_{\mathrm{ht1}}^{\varepsilon_{\Delta t},\mathsf{pR}}$ (Fig.~\ref{ex:cs:pre}) from $\mathcal{M}_{\mathrm{ht1}}$.
We examine the event $E_{\{1,2\}}^{\controller,\varepsilon_{\Delta t},\mathsf{pR}}$, which is for the case where $\hstemp$ satisfies $\phi_{\{1,2\}}^{\controller,\varepsilon_{\Delta t}}(\hstemp) = (\Delta t \leq 5 \Longrightarrow \hstemp \in [30-\Delta t, 30+\Delta t)) \wedge (5 < \Delta t \Longrightarrow \hstemp \in [30-\Delta t, 40-\Delta t])$ (Lines~\ref{line:ex:cs:preStateBegin}--\ref{line:ex:cs:preStateEnd}).

Lines~\ref{line:ex:cs:preStateBegin}--\ref{line:ex:cs:preParamBegin} mean $\pdh \in [30 - (\hstemp - \Delta t), 11]$.
Lines~\ref{line:ex:cs:preStateBegin}--\ref{line:ex:cs:preStateEnd} \& \ref{line:ex:cs:preParamEnd} mean $\pdt \in [30 - \mathrm{max}(30, \hstemp - \Delta t), 40 - \mathrm{min}(40, \hstemp + \Delta t)] \cap [-4,4]$.
Moreover, since the intersection of actions should be nonempty for the feasibility of $E_{\{1,2\}}^{\controller,\varepsilon_{\Delta t},\mathsf{pR}}$ (Lines~\ref{line:ex:cs:preParamIntersection}), $\pdh$ should be equal to $\pdt$.
The existence of such $\pdh$ and $\pdt$
 is equivalent to $\Delta t \leq 2$.
Therefore, an expensive sensor module with $\Delta t \leq 2$ will make the event $E_{\{1,2\}}^{\controller,\varepsilon_{\Delta t},\mathsf{pR}}$ invariant-preserving and feasible.

\begin{figure}[tbp]
 \centering
\begin{lstlisting}
Machine $\mathcal{M}_{\mathrm{ht1}}^{\varepsilon_{\Delta t},\mathsf{pR}}$
  $\ldots$
  Controller event $E_{\{1,2\}}^{\controller,\varepsilon_{\Delta t},\mathsf{pR}}$          /* $\mathsf{ctrl\_heat\_keep\_safe\_eco\_hetero}$ */
    Parameter set $P_{\{1,2\}}^{\controller,\varepsilon_{\Delta t},\mathsf{pR}} = \mathbb{Z} \times \mathbb{Z}$                /* parameter $\pdh$ and $\pdt$ */
    Guard $G_{\{1,2\}}^{\controller,\varepsilon_{\Delta t},\mathsf{pR}} \tuple{\tuple{\tuple{\sturn,\stemp}, \tuple{\hsturn,\hstemp}},\tuple{\pdh,\pdt}} \iff$
      $\;\;\; \forall \tstemp \in [\hstemp-\Delta t, \hstemp+\Delta t] .\, (\tstemp < 30 \vee 30 \leq \tstemp \leq 40)$`\label{line:ex:cs:preStateBegin}`
      $\wedge \; \exists \tstemp \in [\hstemp-\Delta t, \hstemp+\Delta t] .\, (\tstemp < 30)$
      $\wedge \; \exists \tstemp \in [\hstemp-\Delta t, \hstemp+\Delta t] .\, (30 \leq \tstemp \leq 40)$`\label{line:ex:cs:preStateEnd}`
      $\wedge \; \forall \tstemp \in [\hstemp-\Delta t, \hstemp+\Delta t] .\, (\tstemp < 30 \Longrightarrow 30 \leq \tstemp + \pdh \leq 40)$`\label{line:ex:cs:preParamBegin}`
      $\wedge \; \forall \tstemp \in [\hstemp-\Delta t, \hstemp+\Delta t] .\, $ $(30 \leq \tstemp \leq 40 \Longrightarrow 30 \leq \tstemp + \pdt \leq 40 \wedge -4 \leq \pdt \leq 4)$`\label{line:ex:cs:preParamEnd}`
      $\wedge$ $\forall \tstemp \in [\hstemp-\Delta t, \hstemp+\Delta t] .\,$ $(\{\tstemp' \big| \tstemp' = \tstemp + \pdh\} \cap \{\tstemp' \big| \tstemp' = \tstemp + \pdt\}) \neq \emptyset$`\label{line:ex:cs:preParamIntersection}`
    Action $A_{\{1,2\}}^{\controller,\varepsilon_{\Delta t},\mathsf{pR}} \tuple{\tuple{\tuple{\sturn,\stemp}, \tuple{\hsturn,\hstemp}},\tuple{\pdh,\pdt}} = \{\tuple{\tuple{\sturn',\stemp'}, \tuple{\hsturn',\hstemp'}} \,\big|\,$ $\sturn' = \controller \wedge \stemp' = \stemp + \pdh \wedge \stemp' = \stemp + \pdt \wedge \tuple{\sturn',\stemp'} \in \varepsilon_{\Delta t}(\tuple{\hsturn',\hstemp'})\}$`\label{line:ex:cs:preAction}`
  $\ldots$
 \end{lstlisting}
 \caption{The heater model $\mathcal{M}_{\mathrm{ht1}}^{\varepsilon_{\Delta t},\mathsf{pR}}$ produced by action-preserving robustification}
 \label{ex:cs:pre}
\end{figure}

\paragraph{Action-repurposing robustification.}
The model $\mathcal{M}_{\mathrm{ht1}}^{\varepsilon_{\Delta t},\mathsf{rR}}$ (Fig.~\ref{ex:cs:rep}) is generated by the action-repurposing robustification from $\mathcal{M}_{\mathrm{ht1}}$ using the action of $\mathsf{ctrl\_heat}$ event.
We examine the event $E_{\{1,2\}}^{\controller,\varepsilon_{\Delta t},\mathsf{rR}}$ for the $\phi_{\{1,2\}}^{\controller,\varepsilon_{\Delta t}}(\hstemp)$ case again.

Line~\ref{line:ex:cs:repParam} means $\pdh \in [30 + \Delta t - \hstemp, 40 - \Delta t - \hstemp]$.
The existence of such $\pdh$ is equivalent to $30 + \Delta t - \hstemp \leq 40 - \Delta t - \hstemp$, which is also equivalent to $\Delta t \leq 5$.
Thus, we find that we should use a sensor module with $\Delta t = 5$ at least to obtain an invariant-preserving and feasible robustified $E_{\{1,2\}}^{\controller,\varepsilon_{\Delta t},\mathsf{rR}}$.
In this way, the action-repurposing robustification generates a controller that tolerates larger uncertainty at the sacrifice of the compliance with some of original actions (e.g., $\mathcal{M}_{\mathrm{ht1}}^{\varepsilon_{\Delta t},\mathsf{rR}}$ lacks the ecological ``keep\_safe'' functionality).

\begin{figure}[tbp]
 \centering
\begin{lstlisting}
Machine $\mathcal{M}_{\mathrm{ht1}}^{\varepsilon_{\Delta t},\mathsf{rR}}$
  $\ldots$
  Controller event $E_{\{1,2\}}^{\controller,\varepsilon_{\Delta t},\mathsf{rR}}$          /* $\mathsf{ctrl\_heat\_keep\_safe\_eco\_hetero}$ */
    Parameter set $P_{\{1,2\}}^{\controller,\varepsilon_{\Delta t},\mathsf{rR}} = \mathbb{Z} \times \mathbb{Z}$                /* parameter $\pdh$ and $\pdt$ */
    Guard $G_{\{1,2\}}^{\controller,\varepsilon_{\Delta t},\mathsf{rR}} \tuple{\tuple{\tuple{\sturn,\stemp}, \tuple{\hsturn,\hstemp}},\tuple{\pdh,\pdt}} \iff$
      $\;\;\; \forall \tstemp \in [\hstemp-\Delta t, \hstemp+\Delta t] .\, (\tstemp < 30 \vee 30 \leq \tstemp \leq 40)$`\label{line:ex:cs:repStateBegin}`
      $\wedge \; \exists \tstemp \in [\hstemp-\Delta t, \hstemp+\Delta t] .\, (\tstemp < 30)$
      $\wedge \; \exists \tstemp \in [\hstemp-\Delta t, \hstemp+\Delta t] .\, (30 \leq \tstemp \leq 40)$`\label{line:ex:cs:repStateEnd}`
      $\wedge \; \forall \tstemp \in [\hstemp-\Delta t, \hstemp+\Delta t] .\, (30 \leq \tstemp + \pdh \leq 40)$`\label{line:ex:cs:repParam}`
    Action $A_{\{1,2\}}^{\controller,\varepsilon_{\Delta t},\mathsf{rR}} \tuple{\tuple{\tuple{\sturn,\stemp}, \tuple{\hsturn,\hstemp}},\tuple{\pdh,\pdt}} = \{\tuple{\tuple{\sturn',\stemp'}, \tuple{\hsturn',\hstemp'}} \,\big|\,$ $\sturn' = \controller \wedge \stemp' = \stemp + \pdh \wedge \tuple{\sturn',\stemp'} \in \varepsilon_{\Delta t}(\tuple{\hsturn',\hstemp'})\}$
  $\ldots$
 \end{lstlisting}
 \caption{The heater model $\mathcal{M}_{\mathrm{ht1}}^{\varepsilon_{\Delta t},\mathsf{rR}}$ produced by action-repurposing robustification}
 \label{ex:cs:rep}
\end{figure}

\section{Related Work}\label{sec:rw}
The topic of controller robustness to observation noise is a traditional topic in control theory. In this context, the majority of work focuses on robustness with respect to controller stability (e.g., \cite{le2005controller}), rather than arbitrary safety properties. Recent work shows how perceptual uncertainty from visual sensors can be incorporated into the design of a stable controller~\cite{Lipschitz20}.

In the area of controller synthesis from temporal logic specifications, there are approaches for robustifying synthesized controllers by using special interpretations of temporal logic formulas \cite{Fainekos2009,Liu2012,Liu2014}. The basic idea is to contract the regions that must be visited, and inflating those that must be avoided, by $\delta$. These works synthesise hybrid controller implementations with the desired safety property under observation uncertainty; in contrast, in our approach we focus on transforming controller specifications to satisfy the property under observation uncertainty. On the other hand, our current approach is limited to discrete-event and discrete-time systems.

In the context of software systems, Zhang et al.~\cite{Zhang} define robustness as the scope of environmental misbehavior that the system can tolerate without violating its safety property. They find this scope by computing the weakest assumption about the environment that will keep the property, expressed in LTL, satisfied. In contrast, we consider robustness to perceptual uncertainty, and support not only analysis but also automated redesign.

The area of software modeling has a variety of studies on uncertainty, such as combining business rules models and probabilistic relational models for answering probabilistic queries~\cite{10.1007/978-3-319-42019-6_4}, augmenting UML/OCL with new datatypes and operations for modeling and propagation analysis of uncertainty~\cite{Bertoa2020}, and transforming fuzzy UML models into fuzzy description logic knowledge bases for verification~\cite{ZHANG2018134}.
However, to the best of our knowledge, robustifying software models is not proposed.

\section{Conclusion}\label{sec:rwConclusion}

This work provides a workflow to robustify a controller specification against perceptual uncertainty. Since safety properties and action safety are normally specified with respect to the true state of the world, our approach allows designers to first consider the idealized case, and then introduce the perceptual uncertainty as a subsequent step. Our case study demonstrated that our workflow supports the design exploration of the perceptual uncertainty levels that the controller could tolerate. Our methods operate on system specifications expressed in Event-B; however, the ideas of uncertainty injection, and action-preserving and action-repurposing robustification are more general. Specifically, our injection method shows how to introduce perceptual uncertainty into a state machine-based model of an uncertainty-unaware controller.
Our robustification methods take the intersection of applicable actions or calculate parameters that guarantee safety for cases where the controller cannot determine which given actions should be taken due to uncertainty.

In future work, we will extend our methods to improve generality.
For instance, taking probability into account can be promising for extending the application area.
Moreover, we plan to propose a method for systematically relaxing requirements to gain more robustness.

\bibliographystyle{splncs04}
\bibliography{main}
\end{document}